\documentclass[aps,pra,showpacs,twocolumn,amsmath,amssymb]{revtex4-1}
\usepackage{graphicx}
\usepackage{amsmath}
\usepackage{color}
\usepackage{dcolumn}
\usepackage{bm}
\usepackage{booktabs}
\usepackage{subfigure}

\newcommand{\fHe}{$^4$He$^*$}

\newcommand{\mjm}{\mbox{$m$=-1}}
\newcommand{\mjp}{\mbox{$m$=+1}}

\begin{document}

\title{Non-exponential one-body loss in a Bose-Einstein condensate}

\author{S.\,Knoop}
\author{J.\,S.\,Borbely}
\author{R.\,{van Rooij}}
\author{W.\,Vassen}
\affiliation{LaserLaB Vrije Universiteit, De Boelelaan 1081, 1081 HV Amsterdam, the Netherlands}

\date{\today}

\begin{abstract}
We have studied the decay of a Bose-Einstein condensate of metastable helium atoms in an optical dipole trap. In the regime where two- and three-body losses can be neglected we show that the Bose-Einstein condensate and the thermal cloud show fundamentally different decay characteristics. The total number of atoms decays exponentially with time constant $\tau$; however, the thermal cloud decays exponentially with time constant $\frac{4}{3}\tau$ and the condensate decays much faster, and non-exponentially. We show that this behaviour, which should be present for all BECs in thermal equilibrium with a considerable thermal fraction, is due to a transfer of atoms from the condensate to the thermal cloud during its decay.
\end{abstract}


\maketitle

\section{Introduction\label{Introduction}}

Atomic gases can be cooled and trapped to ultracold temperatures and densities where Bose-Einstein condensation (BEC) and Fermi degeneracy can be reached. These trapped gases decay in several ways due to one-body, two-body and three-body collisions. Two- and three-body losses are density dependent and have been extensively studied~\cite{weiner1999,burt1997ccc,soding1998,stamperkurn1998,soding1999,vassen2011cat} and applied in work on atom-atom correlations~\cite{burt1997ccc} and on universal few-body physics~\cite{esry1999,weber2003tbr,kraemer2006efe}.

One-body loss is generally considered trivial as it usually results from collisions between trapped atoms and room-temperature atoms from the background gas in the ultrahigh vacuum chamber that contains the trapped atoms. A background pressure of $\sim$$10^{-11}$~mbar typically leads to a lifetime of atoms in the trap of $\sim$100~s. Other causes of one-body loss are scattering by off-resonant light in a dipole trap and on-resonance excitation by stray laser cooling light. All these effects cause exponential decay of the trapped cloud with a time constant $\tau$, typically in the range of 1 to 100~s. Experiments in ultracold atomic physics measure this time constant by monitoring the number of trapped atoms as a function of time, a procedure commonly performed using absorption imaging of the cloud on a CCD camera. Due to the high densities in a condensate, two- and three-body losses are most important in the first instances of the decay. In studies of decay, it is generally assumed that the decay becomes exponential for long enough times, where two- and three-body losses have become negligible due to the low density that is then reached.

Zin {\it et al.}~\cite{zin2003teo}, however, showed theoretically that the decay of a BEC is expected to be non-exponential when it is in thermal equilibrium with a substantial thermal cloud. The origin of this effect stems from the transfer of atoms from the condensate to the thermal cloud during the decay. This atomic transfer occurs only when thermalization is fast compared to the change of thermodynamic variables during the decay of the trapped cloud. When two-body and three-body decay can be neglected the model predicts that for an overall exponential decay time $\tau$, the condensate decays faster and non-exponentially, while the thermal cloud decays exponentially with a larger time constant, $\frac{4}{3}\tau$. To our knowledge, this has not been demonstrated experimentally.

Enhancement in the decay of a BEC in the presence of a thermal cloud was observed by Tychkov {\it et al.}~\cite{tychkov2006mtb}. In that experiment the decay of a large ($>10^6$ atoms) condensate of helium atoms in the metastable 2 $^3$S$_1$ state (\fHe, lifetime 8000~s) was monitored with and without a thermal cloud containing approximately the same number of atoms. That study, performed in a magnetic trap with atoms in the \mjp~magnetic substate, revealed that the condensate in the presence of a thermal cloud decayed much faster with the cloud than without. However, in that study the decay was studied in the presence of large two- and three-body losses possibly obscuring an effect of atomic transfer; the enhanced decay of the condensate could also be understood from two- and three-body inelastic collisions between condensate and thermal atoms. Also, two- and three-body loss rate constants were not known accurately and both processes were expected to contribute about equally to the decay \cite{tychkov2006mtb,vassen2011cat}. Therefore the rate constants had to be determined from a fit.

Recently, we have accurately determined these two- and three-body loss rate constants over a range of magnetic field values~\cite{borbely2012mfd}. We transferred \fHe~atoms from a magnetic trap into an optical dipole trap and measured, both for atoms in the \mjp~and \mjm~state, the two- and three-body loss rate constants as a function of an applied magnetic field and with a (quasi-)pure BEC. In this paper we extend the previous experiment to partially condensed clouds with a large thermal fraction in order to investigate atomic transfer. To reduce two- and three-body losses we study long time scales and work at small magnetic field using \mjm~atoms that, in contrast to \mjp~atoms, only show three-body losses.

\section{Theory\label{theory}}

\subsection{Theoy of trap loss}

The time evolution of the density of a trapped atomic gas can be described as
\begin{equation}
\label{densityevolution} \dot{n}=-n/\tau-\kappa_2 L_2 n^2-\kappa_3
L_3 n^3.
\end{equation}
The first term on the right takes into account one-body loss, which causes exponential decay with a time constant $\tau$. $L_2$ and $L_3$ are the rate coefficients for two- and three-body loss, respectively, and are defined such that they explicitly include the loss of two and three atoms per loss event. The constants in front of $L_2$ and $L_3$ are $\kappa_2=1/2!$ and $\kappa_3=1/3!$ for a BEC, while $\kappa_2=\kappa_3=1$ for a thermal gas~\cite{kagan1985}.

The rate coefficients are obtained by measuring the number of trapped atoms after a variable hold time. The analysis of this data requires integration of Eq.~\ref{densityevolution} over space. Because a BEC and a thermal cloud have different density distributions, extracting two- and three-body loss rate coefficients from partially condensed samples becomes very difficult, and experiments usually focus on either a pure BEC or a thermal sample.

\subsection{Atomic transfer model}

In Sec.\,2 of their paper, Zin {\it et al.}~\cite{zin2003teo} have derived a simple model for the decay of the thermal and BEC components of a partially condensed cloud in thermal equilibrium below the temperature threshold for Bose-Einstein condensation. The essential assumption of the model is that thermal equilibrium holds during one-body decay of a condensate. Although two- and three-body losses may be incorporated in the model, this complicates the discussion and therefore we here ensure those to be negligible compared to the one-body losses by using low enough densities. If an atom is removed from the thermal cloud by, for instance, a collision with a background atom, then there is a place free in the otherwise saturated thermal distribution, which can be filled by a BEC atom, thus maintaining the size of the thermal cloud at the expense of the BEC. This transfer of atoms from the condensate to the thermal cloud enhances the decay of the condensate and increases the lifetime of the thermal cloud.

For the whole cloud, containing $N$ atoms, decay is exponential with time constant $\tau$. The model then predicts that the thermal cloud will also decay exponentially, however with a larger time constant, while the condensate is expected to decay non-exponentially~\cite{zin2003teo}:
\begin{eqnarray}
\dot{N}&=&-\frac{1}{\tau}N\\
\dot{N_T}&=&-\frac{3}{4\tau}N_T\label{modelequation1}\\
\dot{N_C}&=&-\frac{1}{\tau}\left(N_C+\frac{1}{4}N_T\right)\label{modelequation2}.
\end{eqnarray}
Here $N_C$ and $N_T$ are the number of condensed atoms and number of thermal atoms, respectively, and $N= N_C+N_T$. The equations show that the thermal cloud is expected to decay exponentially with a time constant $\tau'=\frac{4}{3}\tau$, independent of $N_C$, which is assumed to be nonzero. For the BEC a non-exponential decay is expected if an appreciable amount of thermal atoms is present. The model is valid up to the point that there are no BEC atoms left or thermal equilibrium cannot be assumed anymore. Starting with a partially condensed cloud finally leads to the complete depletion of the BEC; for the pure thermal cloud, decay then proceeds with time constant $\tau'=\tau$. Eqs.~\ref{modelequation1} and \ref{modelequation2} are valid when the energy of a condensate atom (which is, apart from a small mean-field contribution, equal to the ground state energy $\epsilon_0$ of the harmonic trap potential) can be neglected compared to the average energy of a thermal atom ($E_T/N_T$)~\cite{zin2003teo}:
\begin{equation}
\label{assumption} \epsilon_0N_T/E_T\ll1.
\end{equation}
In order to observe atomic transfer experimentally, inelastic collisions within the cloud should not play a role, but still a high enough elastic collision rate is necessary to reach fast thermalization. The thermalization rate is given by $\gamma_{\rm th}=\gamma_{\rm coll}/2.7$ \cite{wu1996dso}, with collision rate $\gamma_{\rm coll}=n_{\rm av}\bar{v}\sigma_{\rm el}$. Here, $n_{\rm av}$ is the average density, $\bar{v}=\sqrt{16k_BT/\pi m}$ is the mean relative thermal velocity, and $\sigma_{\rm el}=8\pi a^2$ the elastic cross-section at ultralow temperatures, where $a$=142.0(0.1)$a_0$ for \fHe~\cite{moal2006ado}. The large scattering length  ensures fast thermalization down to a density of $n_{\rm av}=10^{11}$ cm$^{-3}$, which is reached after 40 s in our dipole trap (see Sec.~\ref{Experiment}).

\section{Experiment\label{Experiment}}

We have studied one-body loss in a BEC of \fHe~atoms with a considerable thermal fraction. The experimental setup and measurement procedure have been described earlier~\cite{rooij2011fmi,borbely2012mfd}, here we summarize only the most essential parts. A BEC of about 10$^6$ atoms is prepared in a crossed optical dipole trap at a wavelength of 1557~nm, and all atoms are transferred from the \mjp~magnetic substate to the \mjm~magnetic substate by an RF sweep. We measure the remaining number of trapped atoms after a variable hold time by turning off the trap, causing the atoms to fall and be detected by a micro-channel plate (MCP) detector, which is located 17~cm below the trap center and gives rise to a time-of-flight of approximately 186~ms. From a bimodal fit to the MCP signal, the BEC and thermal fraction are extracted as well as the temperature and the chemical potential of the trapped
gas~\cite{rooij2011fmi}.

\begin{figure}
\includegraphics[width=8.5cm]{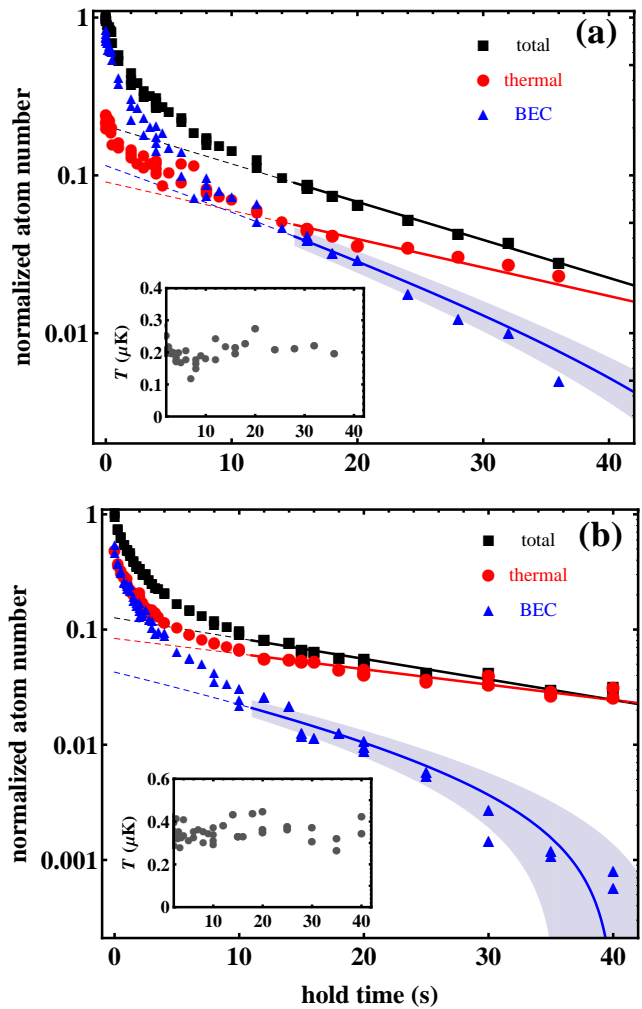}
\caption{(Color online) Lifetime measurements of an ultracold \fHe~gas in the \mjm~state at a magnetic field of 10~G, showing the results of bimodal fits of the MCP signal of the remaining \fHe~atoms, where the insets show the fit temperature. The trap frequencies and the average temperature are (a) $2\pi\times194(1)$~Hz and 0.22(3)~$\mu$K and (b) $2\pi\times245(1)$~Hz and 0.36(5)~$\mu$K, respectively. The solid lines and the shaded area are obtained from fits to the atomic transfer model, as described in the text. \label{lifetimedata}}
\end{figure}

In Fig.~\ref{lifetimedata} we show typical time evolutions of the partially condensed cloud, where the total, thermal and BEC atom number are logarithmically plotted, normalized to the total number of atoms at $t=0$. Fig.~\ref{lifetimedata}(a) shows the decay for a sample with an initial BEC fraction of 80\% (with a thermal fraction at a temperature of 0.22(3)~$\mu$K), while in Fig.~\ref{lifetimedata}(b) the BEC fraction is 50\% (at a temperature of 0.36(5)~$\mu$K). The experiments are performed in a dipole trap with a geometrical mean of the trapping frequency of $2\pi\times194(1)$~Hz and $2\pi\times245(1)$~Hz, respectively. Eq.~\ref{assumption} is easily fulfilled throughout the decay since $\epsilon_0N_T/E_T<0.03$. We also observe that the temperature changes very little during the decay (see the insets in Fig.~\ref{lifetimedata}). At short hold times, the loss is dominated by three-body recombination with a rate constant of $L_3=6.5(0.4)_{\rm stat}(0.6)_{\rm sys}\times10^{-27}$~cm$^6$s$^{-1}$~\cite{borbely2012mfd}. Two-body loss by Penning ionization is strongly suppressed for the spin-stretched states \mjp~and \mjm~\cite{shlyapnikov1994dka}, while for \mjm~the spin-dipole interaction also does not contribute to two-body losses because the atom is in the lowest spin state~\cite{borbely2012mfd}. Fig.~\ref{lifetimedata} clearly shows that, for hold times longer than 10~s, the loss of the total atom number becomes exponential, indicating the regime of one-body loss.
In both measurements the initial density of the BEC is $3\times10^{13}$~cm$^{-3}$. The thermal cloud initially has a density $n_{\rm av}\approx2\times10^{12}$~cm$^{-3}$, which after 40~s has decreased to $n_{\rm av}\approx1\times10^{11}$~cm$^{-3}$. The thermalization rate decreases in that time from $\gamma_{\rm th}\approx50$~s$^{-1}$ to $\gamma_{\rm th}\approx3$~s$^{-1}$, still much larger than the one-body decay rate of approximately 0.05~s$^{-1}$, ensuring sufficiently rapid thermal equilibration.

We observe a faster decay of the BEC fraction than the thermal fraction, eventually leading to a full depletion of the condensate, as expected from the atomic transfer model. Inspection of the decay of the thermal cloud for times longer than 10 s clearly shows that the thermal cloud decays with a larger time constant than the whole cloud. Fitting an exponential decay function to the data at longer hold times yields $\tau=18.0(8)$~s, $\tau'=30.8(2.5)$~s and $\tau'/\tau=1.7(2)$ for the data of Fig.~\ref{lifetimedata}(a), and $\tau=24.4(1.8)$~s, $\tau'=36.2(3.1)$ s and $\tau'/\tau=1.5(2)$ for Fig.~\ref{lifetimedata}(b), where we give $1\sigma$-uncertainties. We attribute the relatively small lifetime to off-resonant scattering of the dipole trap light or resonant excitation by stray light, which could also explain the difference in the obtained lifetimes (our background pressure would limit the lifetime to $\sim$100~s). We conclude that the thermal fraction decays significantly slower than the BEC fraction and that the measured lifetime ratio is in reasonable agreement with the theoretical prediction of $\tau'/\tau=4/3$.

To compare the measurements of the BEC fraction with the atomic transfer model we first fit an exponential decay $N(t)=N_0e^{-t/\tau}$ to the total atom number for long hold times to obtain the overall lifetime $\tau$ as well as $N_0$. Here $N_0$ is the \textit{apparent} atom number at $t$=0 in the absence of three-body loss. In the second step, we fit an exponential decay $N_{T}(t)=N_0(1-f_c)e^{-3t/4\tau}$ to the thermal fraction, with only the \textit{apparent} BEC fraction $f_c$ as a fit parameter. Finally, with the obtained parameters $N_0$, $\tau$ and $f_c$, we numerically solve Eq.~\ref{modelequation2} to obtain $N_C(t)$. The fit results are shown in Fig.~\ref{lifetimedata} as solid lines. The $1\sigma$ errors in the fit parameters are reflected in the band around the theoretical curve for $N_C$. We observe that the atomic transfer model describes the time evolution of the thermal and BEC fraction very well.

\section{Conclusions\label{conclusions}}

Our data provides direct verification of the atomic transfer model. The relatively short lifetime $\tau$~$\approx$~20~s, together with a large scattering length, provide optimal conditions to see this effect. Also, at the relatively small numbers of trapped atoms needed to demonstrate atomic transfer, MCP detection allows better fitting of the BEC fraction compared to absorption imaging. Atomic transfer is most directly visible in the one-body loss regime, but is important in the two- and three-body regime as well, as was theoretically discussed in Ref.~\cite{zin2003teo}. As a final remark we note that one should take care using a BEC for loss measurements. The thermal fraction may be small in the initial stage of BEC decay, but can dramatically increase after long hold times affecting  the decay of the condensate, which will finally become non-exponential.

\begin{acknowledgments}
We thank Jacques Bouma for technical support. This work was financially supported by the Dutch Foundation for Fundamental Research on Matter (FOM). S.\ K.\ acknowledges financial support from the Netherlands Organization for Scientific Research (NWO) via a VIDI grant.
\end{acknowledgments}

\end{document}